\documentclass[journal=mamobx,manuscript=article]{achemso}
\setkeys{acs}{articletitle=true}
\setkeys{acs}{doi=false}

\usepackage[T1]{fontenc}
\usepackage{amsmath}
\usepackage{amssymb}
\usepackage{graphicx}
\usepackage{url}

\SectionNumbersOn
\author{Yi-Xin Liu}
\email{lyx@fudan.edu.cn}
\affiliation{State Key Laboratory of Molecule Engineering of Polymers, Department of
Macromolecular Science, Fudan University, Shanghai 200433, China}

\title{Improved Density Functionals for Predicting Block Copolymer Domain Structure and Phase Behavior}

\begin{document}

\begin{abstract}
Phase-field models offer an efficient alternative to self-consistent field theory (SCFT), but
quantitative predictions of block-copolymer phase behavior remain challenging. We develop a
quadratic-connectivity entropic (QCE) density functional and its adaptive extension (AQCE). QCE
combines quadratic nonlocal connectivity with a nonlinear binary relative-entropy term, while
AQCE adds bounded interfacial stiffness that preserves the homogeneous response. In lamellar
benchmarks, QCE gives the lowest mean profile error among the tested models and AQCE the lowest
mean period error. Compared with the optimized phase-field (OPF) model fitted to SCFT force and
stress data, AQCE reduces strong-segregation period and profile errors by more than fivefold and
sixfold, respectively. AQCE further reproduces the topology of the SCFT phase diagram for AB
diblocks, a more demanding test of free-energy rankings among competing morphologies. Because
architecture enters through ideal-chain correlations, the same nonlinear construction extends to
other A/B architectures. We demonstrate this transfer for symmetric linear ABA lamellae,
accurately predicting their profiles, periods, and ordering threshold.
\end{abstract}

\begin{tocentry}
  \centering
  \includegraphics[width=3.25in]{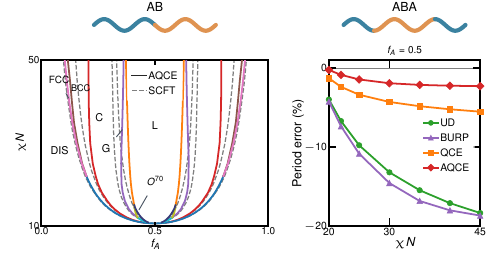}
\end{tocentry}

\section{Introduction}

Block copolymers self-assemble into a rich family of periodic nanostructures whose symmetry and
length scale depend on molecular architecture and segregation
\cite{ref1,ref2,ref3,ref4,ref5,ref6}. Even incompressible AB diblock melts support lamellar
(LAM), cylindrical (CYL), body-centered-cubic spherical (BCC), and double-gyroid (GYR) phases
\cite{ref1,ref2}, while higher-resolution SCFT calculations resolve narrow face-centered-cubic
(FCC) and orthorhombic \(O^{70}\) stability regions \cite{ref7}. Thermodynamic consistency
requires each morphology to be evaluated at its mechanically selected, stress-free unit-cell
period. A quantitative theory must therefore predict density profiles and equilibrium length
scales together with the free-energy differences that determine phase stability.

Self-consistent field theory (SCFT) is the standard quantitative mean-field description of
block-copolymer phase behavior \cite{ref2,ref8,ref9,ref10}. It retains molecular architecture
through forward and backward single-chain propagators, and graph-based formulations now automate
their construction for complex chains \cite{ref11}. This molecular fidelity carries a
computational cost: every nonlinear field iteration requires solving modified diffusion
equations along the chain contour. This contour-resolved calculation becomes particularly costly
for large three-dimensional cells, broad parameter sweeps, and repeated comparisons among
candidate phases. Phase-field, or density-functional, models remove the explicit propagators and
describe the system using species densities or composition fields. A density functional formally
equivalent to Gaussian-chain SCFT can be defined at mean field, but an explicit analytical
approximation is needed to evaluate it without solving the chain-propagator equations
\cite{ref12}.

Several approximations retain selected molecular information while avoiding chain propagators.
The Ohta--Kawasaki (OK) functional combines local mixing thermodynamics with compact gradient
and long-range terms for diblock microphase separation \cite{ref13}. Bohbot-Raviv and Wang (BRW)
combined a nonlinear reference free energy with a quadratic chain-correlation correction and
optimized cell dimensions without prescribing the morphology \cite{ref4}. A complementary route
expands in spatial gradients, as in the density-gradient theory of Tang and Freed for
inhomogeneous polymer blends \cite{ref25}. Uneyama and Doi (UD) developed an
architecture-general construction from ideal-chain correlations and a strong-segregation
extrapolation \cite{ref14}. Subsequent work adapted this framework to reactive polymer blends
\cite{ref26} and applied it to solvent evaporation and reaction-driven membrane assembly
\cite{ref27,ref28}. GPU-accelerated implementations now support large-scale simulations of
blends and solutions \cite{ref29}. The value of these large-scale descriptions depends on how
accurately the underlying functional predicts developed domains and interfaces. To improve
quantitative accuracy, Liu et al. fitted an optimized phase-field (OPF) model to thermodynamic
forces and cell stress from low-dimensional SCFT calculations \cite{ref12}. OPF improves domain
periods and profiles and transfers among morphologies, but its coefficients depend on
composition and segregation strength. The challenge is therefore to combine compactness and
architectural generality with quantitative accuracy across thermodynamic conditions.

The response of a nearly uniform melt constrains the onset of ordering, but developed domains
and interfaces require a nonlinear description. The random-phase approximation (RPA) determines
the quadratic free energy near the homogeneous state, including the spinodal and the preferred
wavelength of weak fluctuations. The higher-order density dependence that governs developed
interfaces and cell stress remains to be specified. Fourth-order RPA expansions provide one
route beyond the Gaussian free energy \cite{ref15}. Uneyama and Doi explicitly recognized this
freedom in extending the weak-segregation response to finite amplitudes \cite{ref14}. In a
recent preprint, Wang et al. applied a nonlocal transformation before the UD strong-segregation
extrapolation and obtained a parameterized family of density functionals \cite{ref16}.
Subsequent reductions of this construction under additional constraints yield functionalized
Cahn--Hilliard descriptions of amphiphilic assembly \cite{ref30}. These precedents show that
preserving weak-response information does not select a unique nonlinear continuation.
Functionals with the same Gaussian composition vertex can therefore predict markedly different
profiles, equilibrium periods, and phase free energies.

We develop the quadratic-connectivity entropic (QCE) density functional to describe these
finite-amplitude density variations while retaining architecture-dependent chain connectivity.
It combines quadratic nonlocal connectivity in the composition field with a binary
relative-entropy local term, with coefficients set by ideal-chain correlations to recover the
reduced RPA-based response. Its adaptive extension (AQCE) allows the interfacial stiffness to
respond to the local density structure through a bounded correction governed by one global
coefficient. Matched QCE/AQCE comparisons isolate the effect of this refinement, while a
full-RPA-matched nonlinear control tests the role of Gaussian response in ordered-state
thermodynamics. We demonstrate marked improvements in density profiles and equilibrium periods
over established density-functional models across broad AB-diblock tests. Beyond these lamellar
benchmarks, AQCE reproduces the topology of the SCFT phase diagram examined for AB diblocks. The
same nonlinear construction also transfers quantitatively to symmetric linear ABA lamellae.

\section{Density Functional Construction}

We first derive the architecture-dependent Gaussian response from ideal-chain correlations for
an incompressible melt of Gaussian block copolymers containing A and B segments
\cite{ref3,ref14}. Incompressibility leaves a single independent composition field, whose
Gaussian vertex fixes the quadratic free-energy curvature \cite{ref1,ref3}. The nonlinear
density dependence governing developed interfaces and cell stress must then be specified. We
examine the UD continuation before constructing the fixed-stiffness QCE functional and its
adaptive extension, AQCE.

\subsection{Gaussian Response of Incompressible Melts}

For a monodisperse incompressible melt of Gaussian block copolymers composed of A and B
segments, with total degree of polymerization \(N\) and common statistical segment length \(b\),
let \(V\) denote the system volume and

\[
\phi(\mathbf r)=\phi_A(\mathbf r),\qquad
\phi_B(\mathbf r)=1-\phi(\mathbf r),\qquad
\psi(\mathbf r)=\phi(\mathbf r)-f, \tag{1}
\]
where \(\phi_A(\mathbf r)\) and \(\phi_B(\mathbf r)\) are the local volume fractions of A and B
segments, respectively, and \(f\) is the overall A-segment fraction. Local incompressibility
ties their fluctuations together, leaving one independent composition fluctuation:

\[
\delta\boldsymbol\phi
=\begin{pmatrix}\delta\phi_A\\\delta\phi_B\end{pmatrix}
=\begin{pmatrix}\psi\\-\psi\end{pmatrix}
=\mathbf v\psi,
\qquad \mathbf v=(1,-1)^\mathsf{T}.
\tag{2}
\]

For a reciprocal-space wavevector \(\mathbf k\) of magnitude \(k=|\mathbf k|\),
\(\boldsymbol{\Omega}(k)\) is the chemistry-resolved ideal-chain intramolecular correlation
matrix. Under the single-chain normalization of Uneyama and Doi,
\(\boldsymbol{\Omega}(0)=N(\mathbf f\otimes\mathbf f)\), with \(\mathbf f=(f,1-f)^\mathsf{T}\)
and \((\mathbf f\otimes\mathbf f)_{\alpha\beta}=f_\alpha f_\beta\) \cite{ref14}. For \(k>0\),
restricting the matrix quadratic form to the allowed fluctuation gives

\[
\widehat{\delta\boldsymbol\phi}^{\dagger}
\boldsymbol{\Omega}(k)^{-1}
\widehat{\delta\boldsymbol\phi}
=|\widehat\psi|^2\,
\mathbf v^\mathsf{T}\boldsymbol{\Omega}(k)^{-1}\mathbf v.
\tag{3}
\]
Thus the ideal-chain second-order composition vertex is

\[
\Gamma_\psi(k)=\mathbf v^\mathsf{T}
\boldsymbol{\Omega}(k)^{-1}\mathbf v.
\tag{4}
\]
This is the incompressible composition projection used in Gaussian-chain RPA
\cite{ref1,ref3,ref14}.

Addition of the Flory--Huggins interaction, parameterized by \(\chi\), gives the collective RPA
composition structure factor \cite{ref1,ref17}

\[
\left[S_\psi^{\rm RPA}(k)\right]^{-1}
=\Gamma_\psi(k)-2\chi.
\tag{5}
\]
The corresponding quadratic free energy is \cite{ref1,ref14}

\[
F^{(2)}=\frac{1}{2V}\sum_{\mathbf k\ne0}
[\Gamma_\psi(k)-2\chi]
|\widehat\psi(\mathbf k)|^2.
\tag{6}
\]
where \(\widehat\psi(\mathbf k)=\int_Vd\mathbf r\, \psi(\mathbf r)e^{-i\mathbf k\cdot\mathbf
r}\), with inverse transform \(\psi(\mathbf r)=V^{-1}\sum_{\mathbf k} \widehat\psi(\mathbf
k)e^{i\mathbf k\cdot\mathbf r}\). Within the Gaussian-chain assumptions, these relations apply
to any A/B block-copolymer architecture \cite{ref3,ref14}: architecture enters through
\(\boldsymbol{\Omega}\) and \(\Gamma_\psi\), while the interaction contributes the shift
\(-2\chi\).

For a linear \(A_fB_{1-f}\) diblock, we use the reduced inverse intramolecular matrix of Uneyama
and Doi \cite{ref14},

\[
\boldsymbol{\Omega}(k)^{-1}\simeq
\frac{\mathbf A}{k^2}+\mathbf C+\frac{b^2k^2}{12}
\begin{pmatrix}
f^{-1}&0\\
0&(1-f)^{-1}
\end{pmatrix}.
\tag{7}
\]
Here \(\mathbf A\) and \(\mathbf C\) denote matrix coefficients: \(\mathbf A\) is the exact
singular coefficient obtained from the small-\(k\) expansion, whereas \(\mathbf C\) is
determined by the Uneyama--Doi intermediate-wavenumber matching prescription \cite{ref14}. Once
\(f\), \(N\), and \(b\) are specified, both matrices are spatially constant and independent of
the local composition field \(\phi(\mathbf r)\). Their scalar projections onto the
incompressible composition direction are

\[
A_\psi=\mathbf v^\mathsf{T}\mathbf A\mathbf v
=\frac{9}{N^2b^2f^2(1-f)^2},\qquad
C_\psi=\mathbf v^\mathsf{T}\mathbf C\mathbf v
=C_{AA}+C_{BB}-2C_{AB}.
\tag{8}
\]
These coefficients give the reduced composition vertex

\[
\Gamma_\psi(k)\simeq\frac{A_\psi}{k^2}+C_\psi+
\frac{b^2k^2}{12f(1-f)}, \tag{9}
\]
where \(C_{AB}=-[4Nf(1-f)]^{-1}\). The diagonal entries of \(\mathbf C\) are fixed by matching
the exact inverse intramolecular matrix at the UD crossover wavevectors:

\[
C_{ii}=[\boldsymbol{\Omega}(k_i)^{-1}]_{ii}
-\frac{A_{ii}}{k_i^2}-\frac{b^2k_i^2}{12f_i},
\qquad k_i^4=\frac{12f_iA_{ii}}{b^2}.
\tag{10}
\]
Here \(i=A,B\), \(f_A=f\), and \(f_B=1-f\). The corresponding diagonal entry of \(\mathbf A\) is
\(A_{ii}=9/(N^2b^2f_i^2)\). Section S1.3 of the Supporting Information details the
intramolecular matrix and crossover construction. The reduced vertex in eq 9 is minimized at the
intrinsic ordering wavevector

\[
k_*^4=\frac{12f(1-f)A_\psi}{b^2}. \tag{11}
\]

Substituting eq 9 into eq 6 gives the reduced quadratic free energy that guides the
finite-amplitude constructions in Sections 2.2 and 2.3.

\subsection{From Gaussian Response to Nonlinear Functionals}

The reduced quadratic free energy specified by eqs 6 and 9 constrains the response near the
uniform melt but does not prescribe a finite-amplitude functional. Uneyama and Doi constructed
such a functional by applying the fixed matrix coefficients \(\mathbf A\) and \(\mathbf C\) in
eq 7 to nonlinear functions of the local block densities \cite{ref14}. For an incompressible AB
diblock, their functional takes the form

\[
\begin{aligned}
F_{\rm UD}[\phi]={}&F_{A,\rm UD}[\phi]+F_{C,\rm UD}[\phi]\\
&+\frac{b^2}{24}\int d\mathbf r\,
\frac{|\nabla\phi|^2}{\phi(1-\phi)}
+\chi\int d\mathbf r\,\phi(1-\phi),
\end{aligned} \tag{12}
\]
where \(F_{A,\rm UD}\) is the square-root-density nonlocal term and \(F_{C,\rm UD}\) contains
logarithmic diagonal terms and a \(\sqrt{\phi(1-\phi)}\) off-diagonal term. Their explicit forms
in the canonical UD normalization \cite{ref14} are collected in Section S1.7. The UD functional
in eq 12 is used for every UD result in Section 3.

The nonlocal, gradient, and contact terms in eq 12 recover \(A_\psi/k^2\),
\(b^2k^2/[12f(1-f)]\), and \(-2\chi\), respectively, in the quadratic expansion about the
uniform melt. The local term behaves differently: its square-root coupling produces a curvature
that differs from \(C_\psi\) in eq 8. Expanding the UD functional about \(\phi=f\) gives the
\(\chi\)-independent composition vertex,

\[
\Gamma_{\rm UD}(k)=\Gamma_\psi(k)+\Delta C_{\rm UD},
\tag{13}
\]
where \(\Gamma_\psi(k)\) is evaluated in the reduced form of eq 9 and

\[
\Delta C_{\rm UD}=-C_{AB}
\left[\frac{1}{f(1-f)}-2\right]>0.
\tag{14}
\]
UD retains the \(k^{-2}\) and \(k^2\) coefficients of eq 9, and hence the preferred
weak-segregation wavevector in eq 11, but its altered local curvature shifts the instability
threshold.

The UD square-root-density form is one admissible extrapolation, not a unique consequence of RPA
\cite{ref14}. Even with a prescribed Gaussian vertex, the higher-order density dependence
remains to be chosen.

\subsection{Quadratic Connectivity and Entropic Closure}

QCE resolves this freedom by retaining quadratic nonlocal connectivity and introducing a
nonlinear entropic local term. The scalar coefficients \(A_\psi\) and \(C_\psi\) obtained in
Section 2.1 specify their required homogeneous curvatures. We choose the inverse-Laplacian
quadratic form in \(\psi\) as the nonlocal continuation because its Fourier multiplier is
exactly \(k^{-2}\), it is positive on the zero-mean composition subspace, and it retains the
standard block-connectivity penalty against macrophase separation \cite{ref13,ref14}. For the
local continuation, we choose the standard binary relative entropy \cite{ref17}

\[
\mathcal E_f(\phi)=
\phi\ln\frac{\phi}{f}+(1-\phi)\ln\frac{1-\phi}{1-f}.
\tag{15}
\]
This convex function vanishes and is stationary at \(\phi=f\), has curvature \(1/[f(1-f)]\)
there, and retains the logarithmic endpoint dependence of ideal mixing. These properties
motivate the local continuation while preserving the required homogeneous curvature.

Combining these choices with the baseline gradient and contact terms gives the QCE functional

\[
\begin{aligned}
F_{\rm QCE}[\phi] ={}&
\frac{A_\psi}{2}\langle\psi,(-\nabla^2)^{-1}\psi\rangle
+M_\psi\int d\mathbf r\,\mathcal E_f(\phi)\\
&+K_{\psi,0}\int d\mathbf r\,
\frac{|\nabla\phi|^2}{\phi(1-\phi)}
+\chi\int d\mathbf r\,[\phi(1-\phi)-f(1-f)].
\end{aligned} \tag{16}
\]
Here \(\langle u,v\rangle\equiv\int_Vd\mathbf r\,u(\mathbf r)v(\mathbf r)\), and
\((-\nabla^2)^{-1}\psi\) denotes the periodic zero-mean solution of \(-\nabla^2u=\psi\). The
operator conventions used in eq 16 are summarized in Section S1.5. Curvature matching fixes the
local coefficient \(M_\psi=C_\psi f(1-f)\) and the segment-normalized gradient coefficient
\(K_{\psi,0}=b^2/24\) inherited from the UD baseline \cite{ref14}. Because this prefactor is
field independent, eq 16 defines fixed-stiffness QCE, with \(1/[\phi(1-\phi)]\) carrying the
composition dependence of the gradient penalty.

Specifically, the four terms in eq 16 continue the \(k^{-2}\), constant, \(k^2\), and
interaction contributions of the reduced quadratic free energy. They represent block
connectivity, local curvature about the homogeneous state, interfacial structure, and contact
interactions, respectively. With \(\phi=f+\psi\), the local and gradient contributions expand as

\[
M_\psi\mathcal E_f(f+\psi)
=\frac{C_\psi}{2}\psi^2+O(\psi^3)
\tag{17}
\]
and

\[
K_{\psi,0}\int d\mathbf r\,
\frac{|\nabla\phi|^2}{\phi(1-\phi)}
=\frac{1}{2V}\sum_{\mathbf k\ne0}
\frac{b^2k^2}{12f(1-f)}|\widehat\psi(\mathbf k)|^2
+O(\psi^3),
\tag{18}
\]
while the contact term contributes \(-2\chi\) to the quadratic response. Together with the
nonlocal term, these expansions give exactly \(\Gamma_\psi(k)-2\chi\), with \(\Gamma_\psi\)
evaluated using eq 9.

These expansions also clarify how QCE differs from the UD continuation. Both UD and QCE use the
incompressible composition restriction in Section 2.1 and retain the same baseline gradient and
contact terms. They continue the \(A_\psi/k^2\) connectivity and local-coefficient contributions
differently at finite amplitude. UD applies the matrix coefficients to square-root and
logarithmic block-density functions, whereas QCE uses an inverse-Laplacian quadratic form in
\(\psi\) and multiplies \(\mathcal E_f\) by \(M_\psi=C_\psi f(1-f)\). These choices recover
\(A_\psi/k^2\) and \(C_\psi\), respectively, in the quadratic expansion of QCE, matching the
reduced vertex in eq 9. A parallel term-by-term derivation of this contrast appears in Section
S1.7.

QCE matches the projected reduced vertex \(\Gamma_\psi\), which raises a complementary question:
is the full RPA vertex sufficient for accurate ordered-state thermodynamics? The quadratic RPA
free energy cannot provide this test because it does not stabilize finite-amplitude ordered
states. To construct a nonlinear control, we use the A/B-antisymmetric logarithmic coordinate

\[
\eta_\psi(\phi)=\frac12\left[
 f\ln\frac{\phi}{f}
 -(1-f)\ln\frac{1-\phi}{1-f}
\right].
\tag{19}
\]
Drawing on the homopolymer logarithmic RPA closure denoted URP, a construction developed by the
author and introduced in Section S1.1 of the Supporting Information, we use the resulting
block-composition URP (BURP) as a nonlinear control. BURP acts on \(\eta_\psi\) with the full
ideal-chain vertex and integrates the resulting force along the straight path from \(f\) to
\(\phi\). Its quadratic response about the homogeneous state is therefore the full RPA inverse
collective composition structure factor \(\Gamma_\psi(k)-2\chi\), with \(\Gamma_\psi\) given by
eq 4. QCE instead uses the reduced form in eq 9 and continues its connectivity, local, gradient,
and contact contributions separately. BURP therefore tests the ordered-state mechanics of a
distinct full-RPA-matched nonlinear continuation. Section S1.6 gives the scalar construction and
verifies its homogeneous curvature.

\subsection{Adaptive Interfacial Stiffness}

Homogeneous curvature matching leaves the finite-amplitude interfacial penalty unresolved.
Strong-segregation theory identifies this penalty, together with chain stretching, as a
determinant of the selected domain scale \cite{ref18}. The homopolymer inverse problem in
Section S1.1 provides a physical clue from density-gradient descriptions of inhomogeneous
polymers \cite{ref17,ref25}. In the homopolymer limits used here, the slow-gradient
approximation (SGA) and the ground-state-dominance (GSD) approximation have the same
density-gradient structure, but their coefficients increase from \(1/6\) to \(1/4\)
\cite{ref17}. Prescribed-density calculations further show that the effective coefficient
depends on profile amplitude and wavenumber. The homopolymer variational-\(K\) (VK1)
construction, developed by the author and introduced in Section S1.1 of the Supporting
Information, captures this dependence through a bounded local coefficient. Transferring this
idea to the block-composition field requires the stiffness to recover \(K_{\psi,0}\) in the
homogeneous limit and remain within the SGA-to-GSD interval, while its adaptive contribution
leaves the Gaussian composition vertex unchanged.

We express the adaptive stiffness in terms of the logarithmic coordinate \(\eta_\psi\) defined
in eq 19. It measures each species' density change relative to its homogeneous fraction in an
A/B-antisymmetric form and satisfies \(\eta_\psi(f)=0\), \(d\eta_\psi/d\phi|_{\phi=f}=1\), and
\(\eta_\psi(1-\phi;1-f)=-\eta_\psi(\phi;f)\). Thus \(\eta_\psi(f+\psi)=\psi+O(\psi^2)\) for weak
variations, while its logarithmic dependence remains sensitive as either species is depleted.

A direct transfer of VK1 gives a nonsmooth block stiffness \(K_{\psi,1}\). AQCE uses a smooth
version, \(K_{\psi,2}\), that retains its sensitivity to local curvature and gradient and its
bounded increase. Section S1.7.3 gives the precursor and smoothing construction. For the final
form, the ordering wavevector \(k_*\) and composition scale \(\eta_*^2=f(1-f)\) provide the
inverse-length and nonzero amplitude scales for the dimensionless invariants

\[
Q=\frac{\nabla^2\eta_\psi}
{k_*^2\sqrt{\eta_\psi^2+\eta_*^2}},
\qquad
P=\frac{|\nabla\eta_\psi|^2}
{k_*^2(\eta_\psi^2+\eta_*^2)}.
\tag{20}
\]
Here \(Q\) measures local curvature and \(P\) measures the squared gradient of the logarithmic
density coordinate. The analytic curvature crossover \(M(Q)=Q^2/\sqrt{1+Q^2}\) is quadratic for
small \(Q\) and approaches \(|Q|\) at large magnitude. Consequently, \(2M(Q)+P\) retains the
relative curvature and gradient weighting of VK1 while removing its absolute-value and max
operations. The hyperbolic tangent then enforces the prescribed bounded increase. With
\(g_2=c_2[2M(Q)+P]\), the AQCE stiffness is

\[
K_{\psi,2}=K_{\psi,0}\left[1+\frac12\tanh g_2\right]. \tag{21}
\]
The differentiable AQCE form is analytic for \(0<\phi<1\), invariant under A/B exchange, and
bounded by \(K_{\psi,0}\le K_{\psi,2}<3K_{\psi,0}/2\). Because \(Q=O(\psi)\),
\(M(Q)=O(\psi^2)\), and \(P=O(\psi^2)\), the adaptive energy begins at fourth order and leaves
the Gaussian composition vertex and spinodal unchanged. The upper bound corresponds to the GSD
stiffness \cite{ref17}. Setting \(K_{\psi,2}=K_{\psi,0}\) recovers the fixed-stiffness QCE in eq
16.

The dimensionless coefficient \(c_2\) controls the sensitivity of the bounded stiffness response
to the local gradient and curvature invariants. A single global value, \(c_2=0.16\), is used
across the tested lamellae, nonlamellar phases, coexistence calculations, and symmetric ABA
transfer. Section S3.1 quantifies the sensitivity of the predictions to this choice.

\subsection{Extension to Other Architectures}

Section 2.1 shows that chain architecture enters QCE and AQCE through the ideal-chain
intramolecular matrix \(\boldsymbol{\Omega}(k)\). Extending the models to another A/B Gaussian
architecture therefore requires recomputing this matrix while retaining the nonlinear closure.
The required block-resolved correlations are evaluated along the chain contour and then grouped
by chemistry into the \(2\times2\) matrix \(\boldsymbol{\Omega}(k)\) \cite{ref3,ref11,ref14}.
Its projection onto the incompressible composition coordinate is the vertex in eq 4. This
construction applies directly to linear and acyclic branched architectures \cite{ref3,ref14}. In
an acyclic chain, each block-pair correlation is evaluated along the unique contour path
connecting the two blocks, as detailed in Section S1.2 \cite{ref11,ref14}. Cyclic architectures,
by contrast, can be treated with graph-based intramolecular correlations, as in related
RPA-based density-functional constructions \cite{ref19}.

We illustrate this procedure with the symmetric linear triblock \(A_{f/2}B_{1-f}A_{f/2}\),
chosen as a convenient example because its symmetry simplifies the chemistry grouping. Label the
terminal A blocks as 1 and 3 and the central B block as 2. Grouping the block-resolved
correlations \(\widetilde\Omega_{ij}\) gives

\[
\begin{aligned}
\Omega_{AA}
&=\widetilde\Omega_{11}+\widetilde\Omega_{33}
+2\widetilde\Omega_{13},\\
\Omega_{BB}&=\widetilde\Omega_{22},\\
\Omega_{AB}&=\widetilde\Omega_{12}+\widetilde\Omega_{32}.
\end{aligned} \tag{22}
\]
The ABA vertex follows by inserting this matrix into the same composition projection. Its
small-\(k\) singularity gives \(A_\psi^{\rm ABA}\), the Uneyama--Doi crossover prescription
gives \(C_\psi^{\rm ABA}\) \cite{ref14}, and the corresponding local coefficient is
\(M_\psi^{\rm ABA}=C_\psi^{\rm ABA}f(1-f)\). The coefficient \(A_\psi^{\rm ABA}\), together with
the universal large-\(k\) gradient term, determines \(k_*^{\rm ABA}\). Replacing the diblock
values of \(A_\psi\), \(M_\psi\), and \(k_*\) in eq 16, the definitions of \(Q\) and \(P\), and
eq 21 completes the ABA extension. The nonlinear form, baseline stiffness \(K_{\psi,0}=b^2/24\),
and global nonlinear stiffness-response coefficient \(c_2=0.16\) remain unchanged. Section S1.4
develops the full ABA matrix construction.

\section{Results and Discussion}

SCFT supplies the quantitative reference for the following tests, with UD and OPF as the
principal literature comparisons and OK as a conventional baseline in the aggregate benchmark.
BURP tests the ordered-state predictions of a full-RPA-matched nonlinear functional, while the
matched QCE/AQCE pair isolates the effect of adaptive stiffness.

\subsection{Spinodal and Weak-Segregation Period}

Before testing ordered states, we first establish which parts of the homogeneous response each
construction preserves. QCE recovers the reduced composition vertex \(\Gamma_\psi(k)\) derived
in Section 2.1, and AQCE follows the same curve because its adaptive contribution begins beyond
quadratic order. BURP retains the exact ideal-chain vertex, whereas the UD functional yields a
different quadratic response (Figure 1a). Adding \(-2\chi\) to each curve gives the complete
second-order composition vertex.

\begin{figure}[H]
  \centering
  \includegraphics[width=\linewidth]{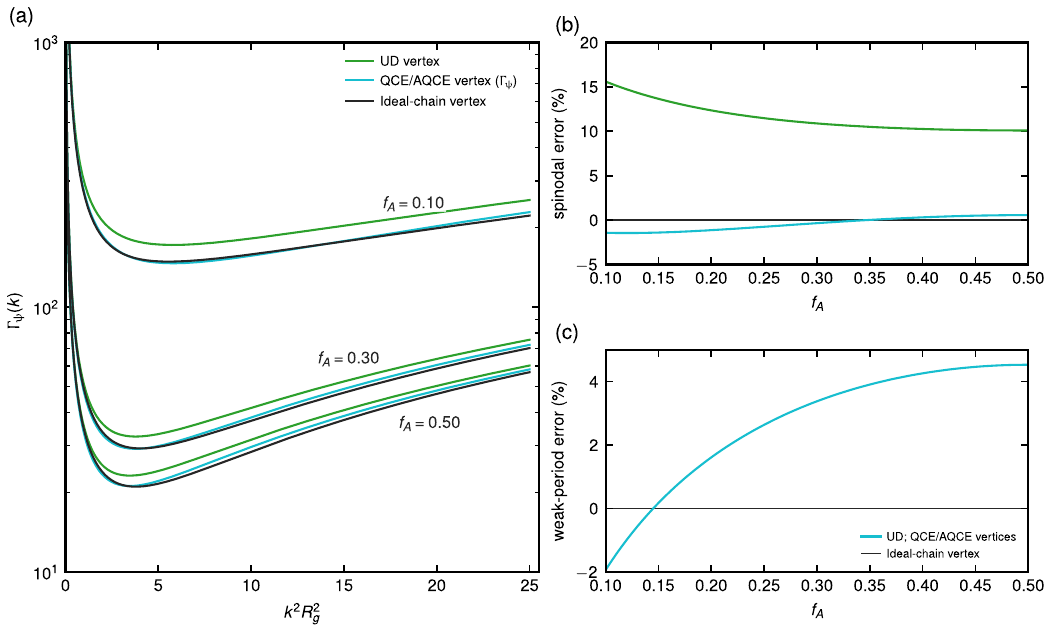}
  \caption{Homogeneous second-order composition vertices and weak-response errors. (a) The \(\chi\)-independent vertices are plotted against \(k^2R_g^2=Nb^2k^2/6\) at \(f_A=0.10,\ 0.30,\) and \(0.50\). Curves denote the exact ideal-chain/BURP vertex, the reduced QCE/AQCE vertex \(\Gamma_\psi\), and the UD vertex \(\Gamma_{\rm UD}\). (b) Relative error in the RPA spinodal and (c) relative weak-period error across composition, using full RPA as the reference.}
  \label{fig:figure1}
\end{figure}

We compare the spinodal and weak-period predictions using their signed relative errors with
respect to full RPA, \(\delta_{\rm sp}=[(\chi N)_{\rm sp}-(\chi N)_{\rm sp}^{\rm RPA}]/ (\chi
N)_{\rm sp}^{\rm RPA}\) and \(\delta_D=(D^*-D_{\rm RPA}^*)/D_{\rm RPA}^*\), where
\(D^*=2\pi/k^*\) is the period of the infinitesimal mode. BURP reproduces both full-RPA
predictions by construction. Matching the local curvature to \(C_\psi\) reduces the absolute
spinodal error of QCE/AQCE relative to UD across the plotted composition range (Figure 1b). The
weak-period prediction, however, depends on the nonlocal-to-gradient coefficient ratio, which is
shared by UD and QCE/AQCE. Consequently, their minimizing wavevectors and weak-period errors
coincide (Figure 1c). Adaptive stiffness enters period selection at finite amplitude, through
its contribution to the ordered-state cell stress.

\subsection{Cell Stress and Period Selection}

To examine how cell stress selects the ordered-state period, we compare the free-energy and
stress branches of UD, BURP, QCE, and AQCE for a symmetric state, \((f_A,\chi N)=(0.50,20)\),
and an asymmetric state, \((0.35,30)\). SCFT provides the reference periods. With
\(K_{\psi,2}=K_{\psi,0}\), QCE provides a matched control that isolates the effect of AQCE's
adaptive stiffness (eq 21). For each model, the density \(\phi_D\) is relaxed at each trial
period \(D\). Following the stress-matching formulation of Liu \textit{et al.} and established
variable-cell field theory \cite{ref12,ref20}, we define the isotropic cell stress as
\(\sigma(D)=\left.\partial(F/V)/\partial\ln D\right|_{\phi_D}\). A stress zero therefore marks a
stationary point of the relaxed free-energy branch. In Figure 2, we subtract each model's
minimum free-energy density and scale both energy and stress by the same positive factor,
preserving the locations of the minima and stress zeros. Root ratios and branch validation are
documented in Section S3.3.

\begin{figure}[H]
  \centering
  \includegraphics[width=\linewidth]{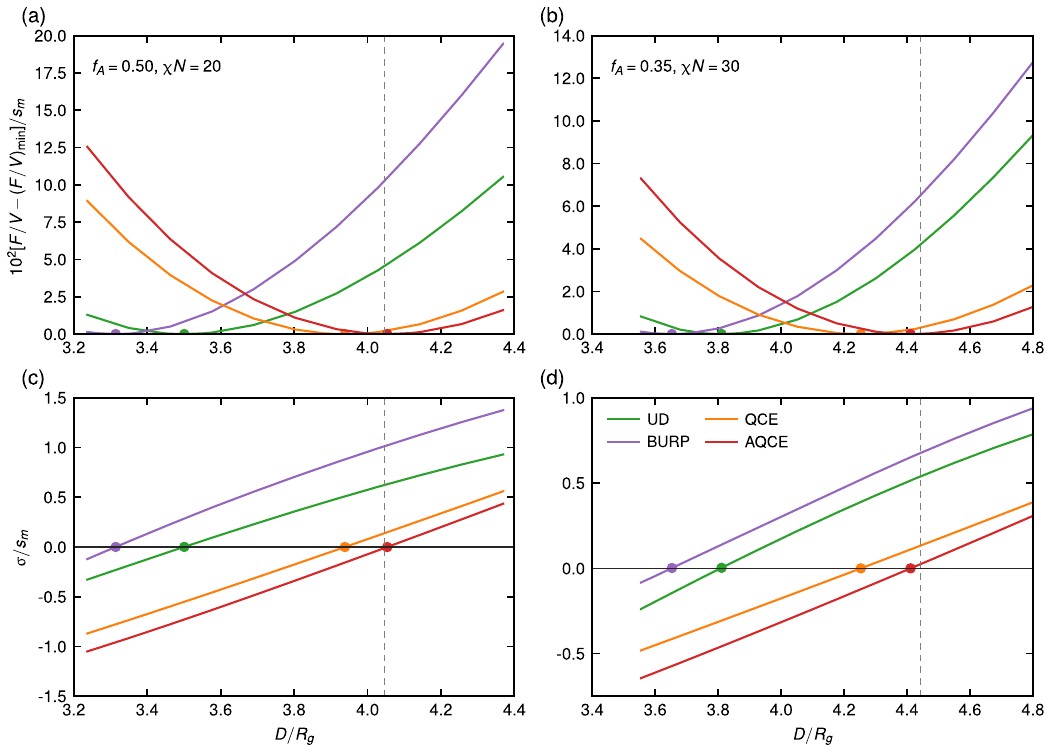}
  \caption{Relaxed free-energy and cell-stress branches. Panels (a) and (c) show the symmetric state \((f_A,\chi N)=(0.50,20)\). Panels (b) and (d) show the asymmetric state \((0.35,30)\). The upper panels give the shifted and scaled free-energy-density branches, and the lower panels give the correspondingly scaled cell stresses. Here, \(s_m=\max\{|F_m(D_{0,m})/V|,1\}\) is the within-model plotting scale. Curves denote UD, BURP, QCE, and AQCE. Filled symbols mark each model's stress-free period. Dashed vertical lines mark the SCFT equilibrium periods.}
  \label{fig:figure2}
\end{figure}

All eight relaxed branches have interior minima (Figures 2a and 2b), and each cell stress
changes sign at the same period (Figures 2c and 2d). The stress zeros expose a clear model
hierarchy. UD and BURP select markedly shorter periods than SCFT in both states, even though the
BURP quadratic response matches the full RPA vertex. The different stress zeros demonstrate the
role of finite-amplitude continuation beyond Gaussian-response matching. In contrast, QCE
substantially reduces the short-period bias, and AQCE moves the zeros closer still. Adaptive
stiffness lowers the absolute period error from 2.655\% to 0.194\% at \((0.50,20)\) and from
4.266\% to 0.711\% at \((0.35,30)\). The invariants \(Q\) and \(P\) activate the bounded
correction where an interface has appreciable curvature and gradient. This mechanism motivates
the broader lamellar benchmark across composition and segregation strength.

\subsection{Lamellar Period and Profile Accuracy}

We next ask whether the cell-mechanical improvement in Section 3.2 persists across composition
and segregation strength. The comparison spans seven compositions, \(f_A=0.20\)--0.50, with
\(\chi N\leq35\) representing weak to intermediate segregation and \(\chi N=36\)--50 probing
strong segregation. Periods and profiles are evaluated in each model's own stress-free cell,
with SCFT as the reference. Across these benchmarks, QCE gives the lowest mean profile error,
while AQCE gives the lowest mean period error and retains high profile accuracy. The relaxation
and acceptance criteria are specified in Sections S2.3, S2.4, and S3.2. We use OPF, whose
condition-specific coefficients are determined from SCFT force and stress data, as the strongest
literature baseline for equilibrium-period accuracy \cite{ref12}. Following Liu \textit{et al.}
\cite{ref12}, the signed relative period error is \(\epsilon=(D_0-D_{0,{\rm SCFT}})/D_{0,{\rm
SCFT}}\). Note that positive \(\epsilon\) denotes a model period longer than the SCFT value.

\begin{figure}[htbp]
  \centering
  \includegraphics[width=\linewidth]{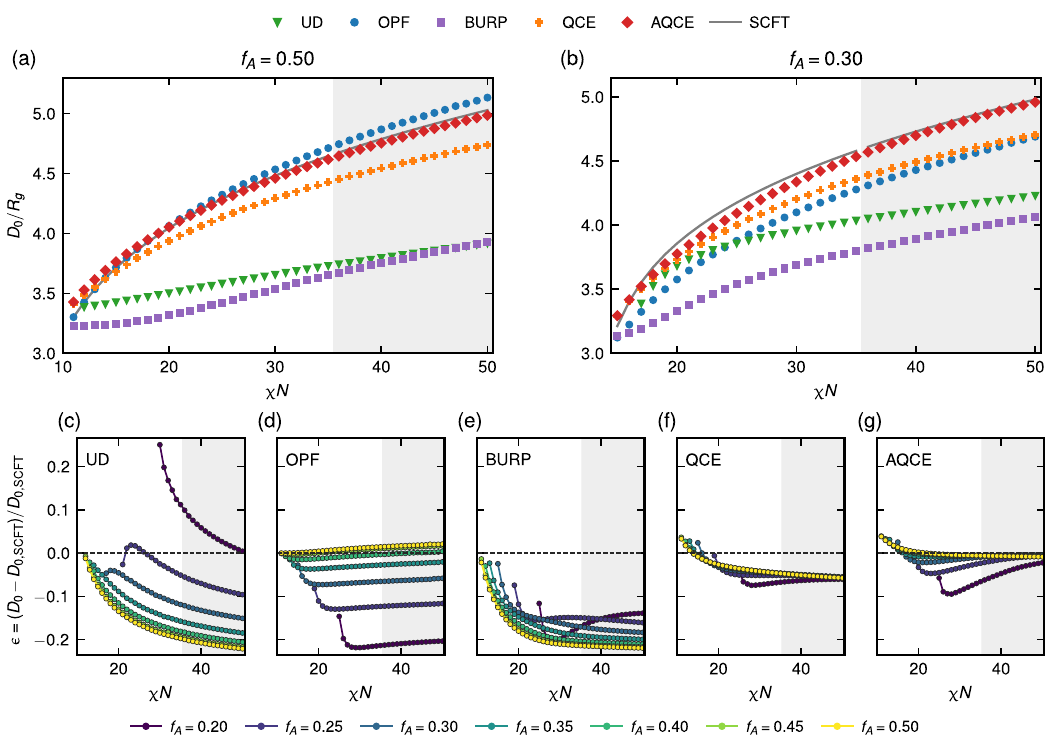}
  \caption{Stress-free lamellar periods and relative errors. Equilibrium period \(D_0/R_g\) for UD, OPF, BURP, QCE, AQCE, and SCFT at (a) \(f_A=0.50\) and (b) \(f_A=0.30\). Panels (c)--(g) give the signed relative period error \(\epsilon\) for UD, OPF, BURP, QCE, and AQCE, respectively, over \(f_A=0.20\)--0.50. Color identifies composition, and shading marks the strong-segregation extension.}
  \label{fig:figure3}
\end{figure}

The advantage of AQCE becomes clearest as segregation strengthens. OPF exhibits its largest
negative errors at low \(f_A\). Along the asymmetric branches, the error passes through a
minimum as \(\chi N\) increases before partially recovering, whereas nearly symmetric branches
develop small positive errors (Figure 3d). In contrast, QCE suppresses much of this composition
dependence, although its negative bias gradually reaches about 5--6\% in strong segregation
(Figure 3f). AQCE further draws the asymmetric branches toward zero and keeps those with
\(f_A\geq0.30\) within about 1\% of SCFT in this regime (Figure 3g). The absolute periods in
Figures 3a and 3b confirm that the benefit of adaptive stiffness grows with segregation
strength.

The benefit of adaptive stiffness depends systematically on composition. Near the upper edge of
the weak-to-intermediate regime, AQCE gives smaller absolute period errors than QCE throughout
\(f_A=0.25\)--0.50. At \(f_A=0.20\), however, QCE remains more accurate over most of the branch
(Figures 3f and 3g). The largest AQCE period error, 9.63\% at \((f_A,\chi N)=(0.20,27)\), also
occurs near ordering at strong asymmetry. The composition-resolved stiffness statistics in
Section S3.3 show a stronger adaptive correction toward symmetric compositions, consistent with
this trend.

Cell relaxation determines the equilibrium profile and period simultaneously. We compare profile
shapes on the normalized coordinate \(s=x/D_0\) using the aligned root-mean-square error
\(\mathrm{RMS}_\phi=[n_s^{-1}\sum_{j=1}^{n_s} (\phi_{A,m}^{\rm align}(s_j)-\phi_{A,\rm
SCFT}(s_j))^2]^{1/2}\). Here \(n_s\) is the number of sampling points, and alignment removes the
arbitrary cell origin. This definition separates profile shape from period error. Sampling and
alignment procedures are given in Section S3.2.

The representative profiles in Figures 4a--d show QCE and AQCE tracking the SCFT profile shapes
for both symmetric and asymmetric compositions near ordering and at stronger segregation. Across
the full maps in Figures 4e--i, QCE produces the lowest aggregate profile error, while AQCE
retains low profile errors alongside its more accurate periods. The comparison of OPF, QCE, and
AQCE in Figures 4f, 4h, and 4i makes this balance particularly clear: OPF profile errors grow
with segregation strength, especially at asymmetric compositions, whereas QCE and AQCE maintain
substantially smaller errors in that regime.

\begin{figure}[htbp]
  \centering
  \includegraphics[width=\linewidth]{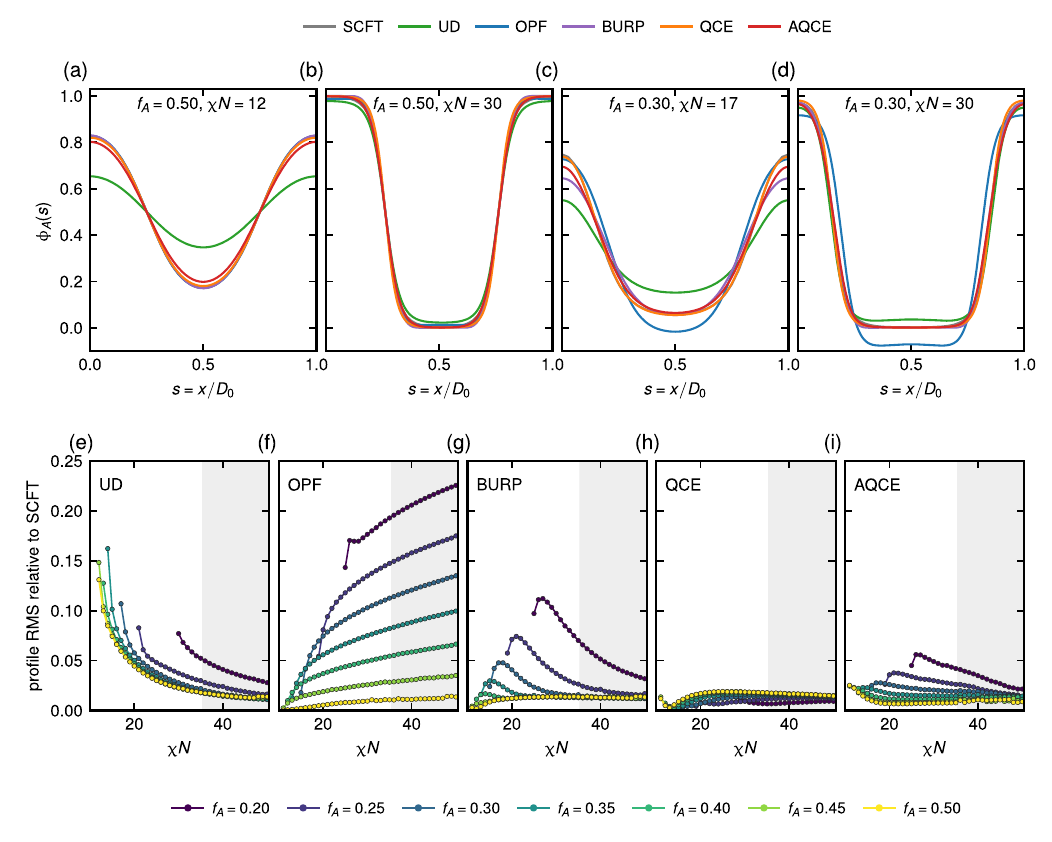}
  \caption{Stress-free lamellar profiles and profile errors. Panels (a)--(d) compare SCFT, UD, OPF, BURP, QCE, and AQCE profiles at \((f_A,\chi N)=(0.50,12)\), \((0.50,30)\), \((0.30,17)\), and \((0.30,30)\), respectively. The profiles are shown in each model's stress-free cell using \(s=x/D_0\) and periodic alignment. Panels (e)--(i) give the aligned profile RMS relative to SCFT for UD, OPF, BURP, QCE, and AQCE, respectively, over \(f_A=0.20\)--0.50. Shading marks the strong-segregation extension.}
  \label{fig:figure4}
\end{figure}

Aggregation across the two segregation regimes sharpens the complementary roles of QCE and AQCE
(Figure 5). Within each regime, the bars are unweighted means over the common set of conditions
satisfying the validation criteria for every compared model. OK provides the conventional
reduced baseline, and OPF has the lowest mean period error among the literature models in both
regimes. Relative to OPF in weak to intermediate segregation, QCE reduces the mean absolute
period error from 4.31\% to 3.62\% and the mean profile RMS from 0.058 to 0.013. AQCE reduces
the period error further to 1.46\%, with a profile RMS of 0.017.

Strong segregation separates the models more sharply. Relative to OPF, QCE reduces the profile
RMS by more than sevenfold, while AQCE reduces the period error by more than fivefold and the
profile RMS by more than sixfold. AQCE also lowers the QCE period error by more than fourfold.
These gains arise from complementary parts of the construction. The QCE results identify the
base nonlinear closure as the principal source of improved profile fidelity, while the AQCE
results show how bounded adaptive stiffness further corrects period selection.

\begin{figure}[htbp]
  \centering
  \includegraphics[width=\linewidth]{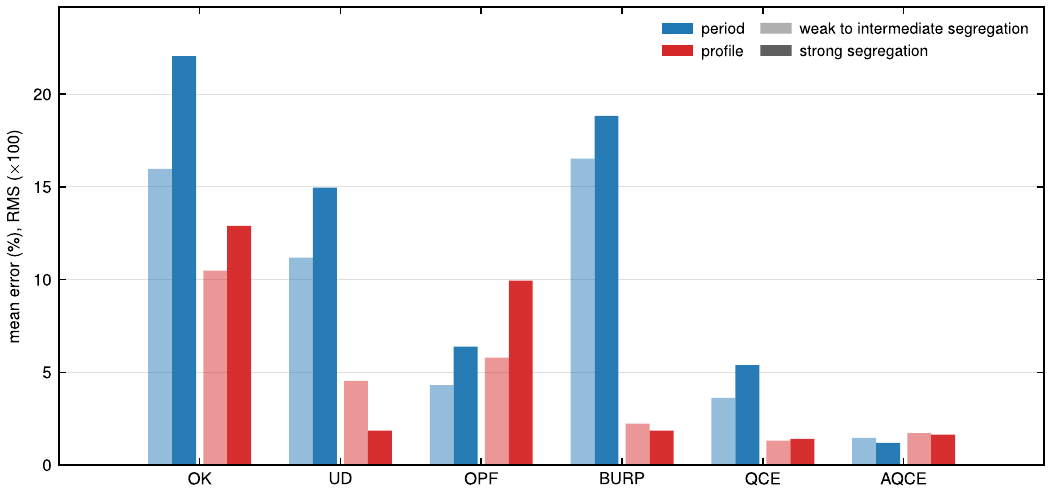}
  \caption{Aggregate stress-free lamellar accuracy. Mean absolute relative period error (blue) and \(100\) times the aligned \(\phi_A\)-profile RMS relative to SCFT (red) for OK, UD, OPF, BURP, QCE, and AQCE. Lighter bars denote the common weak-to-intermediate-segregation benchmark set, and darker bars denote strong segregation.}
  \label{fig:figure5}
\end{figure}

\subsection{Diblock Phase Diagram Topology}

The AB-diblock phase diagram provides a more stringent test than lamellar profiles or periods
alone because several competing morphologies must be ranked consistently after cell relaxation.
AQCE recovers DIS and the six ordered phases FCC, BCC, CYL, GYR, \(O^{70}\), and LAM, with the
same connectivity of stability regions as in the SCFT phase diagram reported by Li and Liu
\cite{ref7} (Figure 6). Boundary construction and numerical procedures are detailed in Section
S4.

\begin{figure}[htbp]
  \centering
  \includegraphics[width=\linewidth]{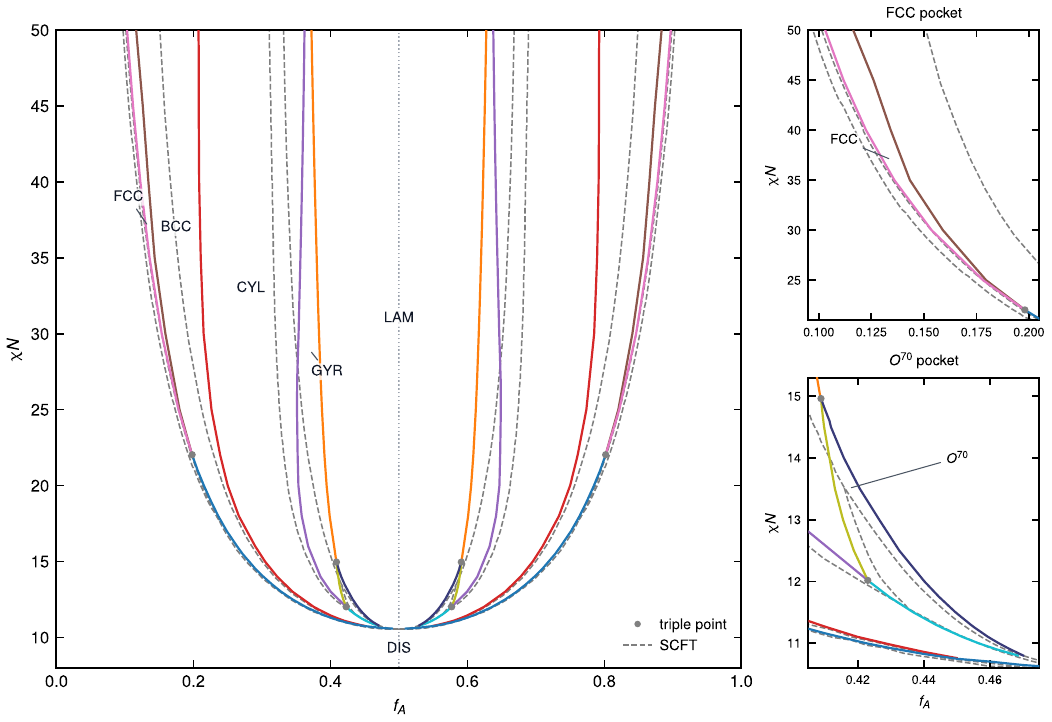}
  \caption{AQCE diblock phase diagram. Solid curves show AQCE boundaries for \(f_A\le0.5\), mirrored by A/B symmetry. Dashed curves show SCFT boundaries \cite{ref7}, and gray disks mark estimated triple points. Labels identify FCC and BCC spheres, cylinders (CYL), gyroid (GYR), the orthorhombic network (\(O^{70}\)), lamellae (LAM), and disorder (DIS). Right-hand panels enlarge the FCC and \(O^{70}\) regions using the same curves and triple points.}
  \label{fig:figure6}
\end{figure}

The persistence of the GYR region at strong segregation distinguishes AQCE from UD. In the UD
phase diagram, the pocket narrows and closes between \(\chi N=40\) and 50, leaving a direct
CYL/LAM boundary at higher segregation \cite{ref14}. In contrast, SCFT retains a stable GYR
phase in strong segregation \cite{ref7,ref22,ref23}. AQCE reproduces this persistence: both its
GYR/CYL and GYR/LAM boundaries extend through \(\chi N=50\), preserving the GYR channel between
CYL and LAM.

The narrow FCC and \(O^{70}\) regions further test the small free-energy differences between
competing morphologies. AQCE recovers the FCC pocket between DIS and BCC and the \(O^{70}\)
channel connecting CYL, GYR, and LAM. This agreement demonstrates that AQCE captures the
relative stability of structurally distinct ordered phases, extending the validation beyond
lamellar periods and profiles.

\subsection{Transfer to ABA Triblocks}

Following the construction in Section 2.5, we test architectural transfer by replacing the
diblock ideal-chain coefficients with those of symmetric linear ABA triblocks, while retaining
the same nonlinear closure and adaptive stiffness.

\begin{figure}[htbp]
  \centering
  \includegraphics[width=\linewidth]{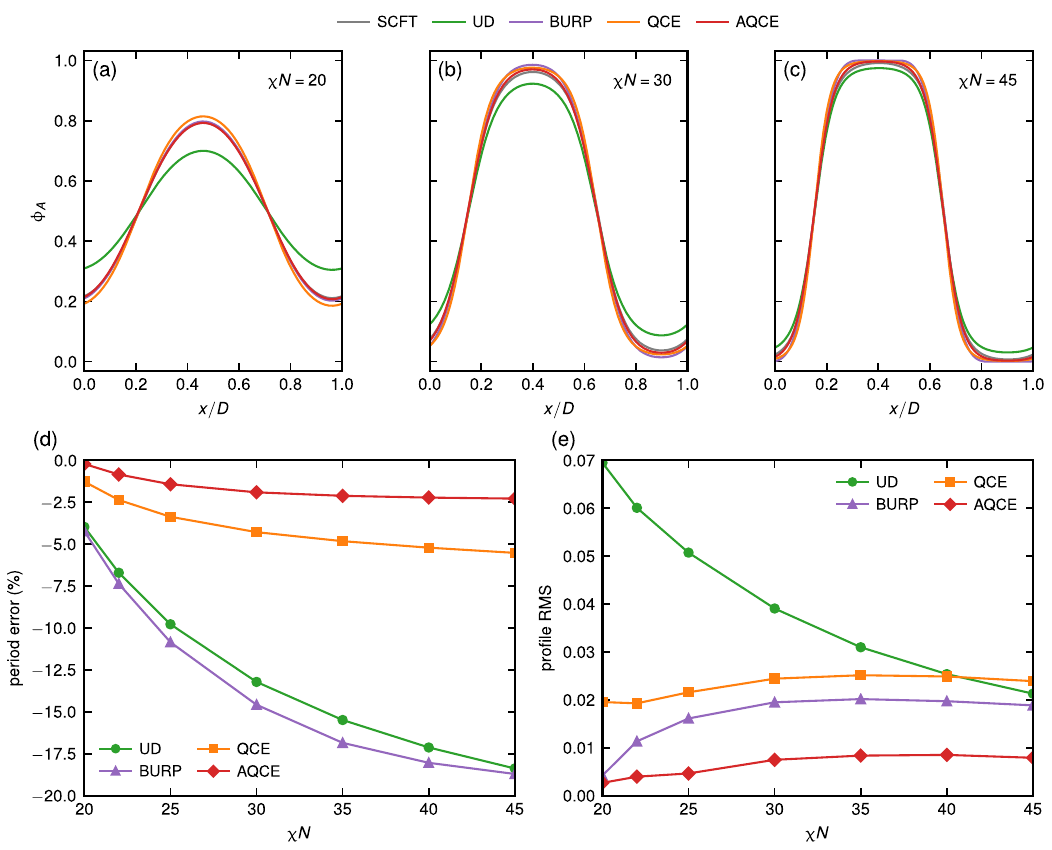}
  \caption{Transfer to symmetric linear ABA lamellae. Panels (a)--(c) compare profiles for SCFT, UD, BURP, QCE, and AQCE at \(f_A=0.50\) and \(\chi N=20\), 30, and 45. Panels (d) and (e) give the stress-free-period error and aligned profile RMS relative to SCFT over the seven transfer conditions.}
  \label{fig:figure7}
\end{figure}

AQCE closely follows the SCFT profiles as the interfaces sharpen from weak to strong segregation
(Figures 7a--c). UD and BURP develop pronounced short-period biases as segregation strengthens,
whereas QCE and AQCE substantially reduce this bias (Figure 7d). For the matched QCE/AQCE pair,
adaptive stiffness improves both period and profile accuracy across the tested ABA states
(Figures 7d and 7e), in contrast to the aggregate tradeoff in the diblock benchmark. The mean
absolute period error decreases from 3.83\% for QCE to 1.57\% for AQCE, and the mean profile RMS
from 0.023 to 0.006. AQCE has the smaller absolute period error at every tested interaction
strength. Detailed profile metrics are reported in Section S5.

\begin{figure}[H]
  \centering
  \includegraphics[width=0.75\linewidth]{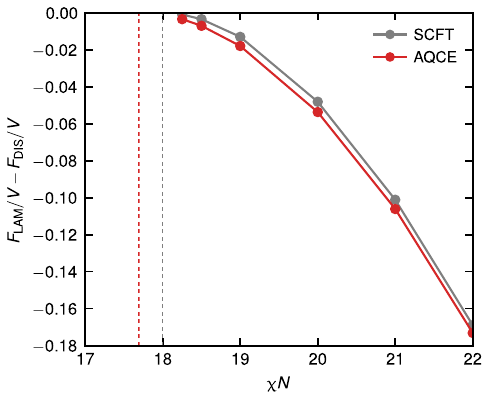}
  \caption{LAM--DIS bifurcation for symmetric linear ABA triblocks. Free-energy difference between the relaxed LAM and DIS branches for AQCE and SCFT. Symbols mark calculated ordered branches, and vertical dotted lines show the corresponding Gaussian bifurcations.}
  \label{fig:figure8}
\end{figure}

Beyond the ordered lamellae, the LAM--DIS threshold tests the transferred functional as order
first emerges. The AQCE L/DIS bifurcation occurs at \(\chi N=17.69\), compared with 18.00 for
SCFT (Figure 8). Above their respective bifurcations, both models predict a similar decrease in
the free-energy difference between the LAM and DIS branches as segregation strengthens.
Together, these results demonstrate quantitative transfer of the same nonlinear construction
from AB diblocks to symmetric linear ABA lamellae.

\section{Conclusions}

We have developed nonlinear density functionals that combine quadratic nonlocal connectivity
with a binary relative-entropy local term. QCE retains the reduced Gaussian composition
response, while AQCE introduces bounded adaptive stiffness to improve the description of
developed interfaces and cell stress. The comparison with the full-RPA-matched BURP control
demonstrates the importance of the nonlinear continuation in determining ordered-state
thermodynamics.

The AB-diblock benchmarks reveal complementary strengths of the two models. QCE gives the lowest
aggregate profile error among the tested models, while AQCE gives the lowest aggregate period
error. These results connect improved profile fidelity to the base nonlinear closure and
improved period selection to the adaptive interfacial correction. The advantage over the
SCFT-informed OPF model becomes especially pronounced in strong segregation, where AQCE reduces
the mean period error by more than fivefold and the profile RMS by more than sixfold.

Beyond lamellar accuracy, AQCE reproduces the topology of the SCFT phase diagram examined for AB
diblocks, including the persistent GYR channel and the narrow FCC and \(O^{70}\) stability
regions. This agreement demonstrates that the functional also captures the relative stability of
competing morphologies. Furthermore, replacing the ideal-chain coefficients while retaining the
nonlinear closure and adaptive stiffness yields accurate profiles, periods, and ordering
thresholds for symmetric linear ABA lamellae. This transfer motivates applications to other A/B
sequences and acyclic branched architectures.

The improved thermodynamic description provides a basis for extending QCE and AQCE to
nonequilibrium self-assembly. Recent UD-based studies of reactive polymer blends and
reaction-driven membrane assembly offer concrete directions for this extension
\cite{ref26,ref28}. Future work can combine multicomponent thermodynamics with conserved
transport and reaction kinetics to investigate how the improved description of interfaces and
phase stability influences assembly pathways.

\begin{acknowledgement}
This work was supported by the National Natural Science Foundation of China (Grants No. 21873021).
\end{acknowledgement}

\section*{Conflict of Interests}

The authors declare no conflict of interest.

\section*{Associated Content}

\textbf{Supporting Information.} Theoretical derivations and analytic derivatives; numerical
methods and convergence criteria; coefficient estimation and lamellar benchmarks; phase-boundary
construction and SCFT reference comparison; and symmetric ABA validation (PDF).

\noindent \textbf{Data and Code Availability.} Materials supporting this work are available at
\url{https://github.com/PolyorderLab/qce-aqce-paper}.

\section*{Author Information}

\textbf{Corresponding Author.} Yi-Xin Liu, State Key Laboratory of Molecule Engineering of
Polymers, Department of Macromolecular Science, Fudan University, Shanghai 200433, China;
lyx@fudan.edu.cn.

\nocite{ref24}
\bibliography{references}

\providecommand{\latin}[1]{#1}
\makeatletter
\providecommand{\doi}
  {\begingroup\let\do\@makeother\dospecials
  \catcode`\{=1 \catcode`\}=2 \doi@aux}
\providecommand{\doi@aux}[1]{\endgroup\texttt{#1}}
\makeatother
\providecommand*\mcitethebibliography{\thebibliography}
\csname @ifundefined\endcsname{endmcitethebibliography}  {\let\endmcitethebibliography\endthebibliography}{}
\begin{mcitethebibliography}{30}
\providecommand*\natexlab[1]{#1}
\providecommand*\mciteSetBstSublistMode[1]{}
\providecommand*\mciteSetBstMaxWidthForm[2]{}
\providecommand*\mciteBstWouldAddEndPuncttrue
  {\def\EndOfBibitem{\unskip.}}
\providecommand*\mciteBstWouldAddEndPunctfalse
  {\let\EndOfBibitem\relax}
\providecommand*\mciteSetBstMidEndSepPunct[3]{}
\providecommand*\mciteSetBstSublistLabelBeginEnd[3]{}
\providecommand*\EndOfBibitem{}
\mciteSetBstSublistMode{f}
\mciteSetBstMaxWidthForm{subitem}{(\alph{mcitesubitemcount})}
\mciteSetBstSublistLabelBeginEnd
  {\mcitemaxwidthsubitemform\space}
  {\relax}
  {\relax}

\bibitem[Leibler(1980)]{ref1}
Leibler,~L. Theory of Microphase Separation in Block Copolymers. \emph{Macromolecules} \textbf{1980}, \emph{13}, 1602--1617\relax
\mciteBstWouldAddEndPuncttrue
\mciteSetBstMidEndSepPunct{\mcitedefaultmidpunct}
{\mcitedefaultendpunct}{\mcitedefaultseppunct}\relax
\EndOfBibitem
\bibitem[Matsen and Schick(1994)Matsen, and Schick]{ref2}
Matsen,~M.~W.; Schick,~M. Stable and Unstable Phases of a Diblock Copolymer Melt. \emph{Phys. Rev. Lett.} \textbf{1994}, \emph{72}, 2660--2663\relax
\mciteBstWouldAddEndPuncttrue
\mciteSetBstMidEndSepPunct{\mcitedefaultmidpunct}
{\mcitedefaultendpunct}{\mcitedefaultseppunct}\relax
\EndOfBibitem
\bibitem[Morozov and Fraaije(2001)Morozov, and Fraaije]{ref3}
Morozov,~A.~N.; Fraaije,~J. G. E.~M. Phase Behavior of Block Copolymer Melts with Arbitrary Architecture. \emph{J. Chem. Phys.} \textbf{2001}, \emph{114}, 2452--2465\relax
\mciteBstWouldAddEndPuncttrue
\mciteSetBstMidEndSepPunct{\mcitedefaultmidpunct}
{\mcitedefaultendpunct}{\mcitedefaultseppunct}\relax
\EndOfBibitem
\bibitem[Bohbot-Raviv and Wang(2000)Bohbot-Raviv, and Wang]{ref4}
Bohbot-Raviv,~Y.; Wang,~Z.-G. Discovering New Ordered Phases of Block Copolymers. \emph{Phys. Rev. Lett.} \textbf{2000}, \emph{85}, 3428--3431\relax
\mciteBstWouldAddEndPuncttrue
\mciteSetBstMidEndSepPunct{\mcitedefaultmidpunct}
{\mcitedefaultendpunct}{\mcitedefaultseppunct}\relax
\EndOfBibitem
\bibitem[Xie and Shi(2024)Xie, and Shi]{ref5}
Xie,~J.; Shi,~A.-C. Phase Behavior of Triblock Copolymer and Homopolymer Blends: Effect of Copolymer Topology. \emph{Phys. Rev. Mater.} \textbf{2024}, \emph{8}, 015601\relax
\mciteBstWouldAddEndPuncttrue
\mciteSetBstMidEndSepPunct{\mcitedefaultmidpunct}
{\mcitedefaultendpunct}{\mcitedefaultseppunct}\relax
\EndOfBibitem
\bibitem[He \latin{et~al.}(2002)He, Huang, Liang, and Pan]{ref6}
He,~X.; Huang,~L.; Liang,~H.; Pan,~C. Self-Assembly of Star Block Copolymers by Dynamic Density Functional Theory. \emph{J. Chem. Phys.} \textbf{2002}, \emph{116}, 10508--10513\relax
\mciteBstWouldAddEndPuncttrue
\mciteSetBstMidEndSepPunct{\mcitedefaultmidpunct}
{\mcitedefaultendpunct}{\mcitedefaultseppunct}\relax
\EndOfBibitem
\bibitem[Li and Liu(2021)Li, and Liu]{ref7}
Li,~W.; Liu,~Y.-X. Simplicity in Mean-Field Phase Behavior of Two-Component Miktoarm Star Copolymers. \emph{J. Chem. Phys.} \textbf{2021}, \emph{154}, 014903\relax
\mciteBstWouldAddEndPuncttrue
\mciteSetBstMidEndSepPunct{\mcitedefaultmidpunct}
{\mcitedefaultendpunct}{\mcitedefaultseppunct}\relax
\EndOfBibitem
\bibitem[Matsen and Bates(1996)Matsen, and Bates]{ref8}
Matsen,~M.~W.; Bates,~F.~S. Unifying Weak- and Strong-Segregation Block Copolymer Theories. \emph{Macromolecules} \textbf{1996}, \emph{29}, 1091--1098\relax
\mciteBstWouldAddEndPuncttrue
\mciteSetBstMidEndSepPunct{\mcitedefaultmidpunct}
{\mcitedefaultendpunct}{\mcitedefaultseppunct}\relax
\EndOfBibitem
\bibitem[Matsen(2002)]{ref9}
Matsen,~M.~W. The Standard Gaussian Model for Block Copolymer Melts. \emph{J. Phys.: Condens. Matter} \textbf{2002}, \emph{14}, R21--R47\relax
\mciteBstWouldAddEndPuncttrue
\mciteSetBstMidEndSepPunct{\mcitedefaultmidpunct}
{\mcitedefaultendpunct}{\mcitedefaultseppunct}\relax
\EndOfBibitem
\bibitem[Arora \latin{et~al.}(2016)Arora, Qin, Morse, Delaney, Fredrickson, Bates, and Dorfman]{ref10}
Arora,~A.; Qin,~J.; Morse,~D.~C.; Delaney,~K.~T.; Fredrickson,~G.~H.; Bates,~F.~S.; Dorfman,~K.~D. Broadly Accessible Self-Consistent Field Theory for Block Polymer Materials Discovery. \emph{Macromolecules} \textbf{2016}, \emph{49}, 4675--4690\relax
\mciteBstWouldAddEndPuncttrue
\mciteSetBstMidEndSepPunct{\mcitedefaultmidpunct}
{\mcitedefaultendpunct}{\mcitedefaultseppunct}\relax
\EndOfBibitem
\bibitem[Zhang \latin{et~al.}(2024)Zhang, Huang, and Liu]{ref11}
Zhang,~Y.; Huang,~W.; Liu,~Y.-X. Automated Chain Architecture Screening for Discovery of Block Copolymer Assembly with Graph Enhanced Self-Consistent Field Theory. \emph{Commun. Mater.} \textbf{2024}, \emph{5}, 266\relax
\mciteBstWouldAddEndPuncttrue
\mciteSetBstMidEndSepPunct{\mcitedefaultmidpunct}
{\mcitedefaultendpunct}{\mcitedefaultseppunct}\relax
\EndOfBibitem
\bibitem[Liu \latin{et~al.}(2019)Liu, Garc{\'i}a-Cervera, Delaney, and Fredrickson]{ref12}
Liu,~J.~V.; Garc{\'i}a-Cervera,~C.~J.; Delaney,~K.~T.; Fredrickson,~G.~H. Optimized Phase Field Model for Diblock Copolymer Melts. \emph{Macromolecules} \textbf{2019}, \emph{52}, 2878--2888\relax
\mciteBstWouldAddEndPuncttrue
\mciteSetBstMidEndSepPunct{\mcitedefaultmidpunct}
{\mcitedefaultendpunct}{\mcitedefaultseppunct}\relax
\EndOfBibitem
\bibitem[Ohta and Kawasaki(1986)Ohta, and Kawasaki]{ref13}
Ohta,~T.; Kawasaki,~K. Equilibrium Morphology of Block Copolymer Melts. \emph{Macromolecules} \textbf{1986}, \emph{19}, 2621--2632\relax
\mciteBstWouldAddEndPuncttrue
\mciteSetBstMidEndSepPunct{\mcitedefaultmidpunct}
{\mcitedefaultendpunct}{\mcitedefaultseppunct}\relax
\EndOfBibitem
\bibitem[Tang and Freed(1991)Tang, and Freed]{ref25}
Tang,~H.; Freed,~K.~F. Free Energy Functional Expansion for Inhomogeneous Polymer Blends. \emph{J. Chem. Phys.} \textbf{1991}, \emph{94}, 1572--1583\relax
\mciteBstWouldAddEndPuncttrue
\mciteSetBstMidEndSepPunct{\mcitedefaultmidpunct}
{\mcitedefaultendpunct}{\mcitedefaultseppunct}\relax
\EndOfBibitem
\bibitem[Uneyama and Doi(2005)Uneyama, and Doi]{ref14}
Uneyama,~T.; Doi,~M. Density Functional Theory for Block Copolymer Melts and Blends. \emph{Macromolecules} \textbf{2005}, \emph{38}, 196--205\relax
\mciteBstWouldAddEndPuncttrue
\mciteSetBstMidEndSepPunct{\mcitedefaultmidpunct}
{\mcitedefaultendpunct}{\mcitedefaultseppunct}\relax
\EndOfBibitem
\bibitem[Tikekar \latin{et~al.}(2022)Tikekar, Delaney, Villet, Tree, and Fredrickson]{ref26}
Tikekar,~M.~D.; Delaney,~K.~T.; Villet,~M.~C.; Tree,~D.~R.; Fredrickson,~G.~H. A Phase Field Model for Dynamic Simulations of Reactive Blending of Polymers. \emph{Soft Matter} \textbf{2022}, \emph{18}, 877--893\relax
\mciteBstWouldAddEndPuncttrue
\mciteSetBstMidEndSepPunct{\mcitedefaultmidpunct}
{\mcitedefaultendpunct}{\mcitedefaultseppunct}\relax
\EndOfBibitem
\bibitem[Dreyer \latin{et~al.}(2022)Dreyer, Ibbeken, Schneider, Blagojevic, Radjabian, Abetz, and M{\"u}ller]{ref27}
Dreyer,~O.; Ibbeken,~G.; Schneider,~L.; Blagojevic,~N.; Radjabian,~M.; Abetz,~V.; M{\"u}ller,~M. Simulation of Solvent Evaporation from a Diblock Copolymer Film: Orientation of the Cylindrical Mesophase. \emph{Macromolecules} \textbf{2022}, \emph{55}, 7564--7582\relax
\mciteBstWouldAddEndPuncttrue
\mciteSetBstMidEndSepPunct{\mcitedefaultmidpunct}
{\mcitedefaultendpunct}{\mcitedefaultseppunct}\relax
\EndOfBibitem
\bibitem[H{\"a}fner and M{\"u}ller(2023)H{\"a}fner, and M{\"u}ller]{ref28}
H{\"a}fner,~G.; M{\"u}ller,~M. Reaction-Driven Assembly: Controlling Changes in Membrane Topology by Reaction Cycles. \emph{Soft Matter} \textbf{2023}, \emph{19}, 7281--7292\relax
\mciteBstWouldAddEndPuncttrue
\mciteSetBstMidEndSepPunct{\mcitedefaultmidpunct}
{\mcitedefaultendpunct}{\mcitedefaultseppunct}\relax
\EndOfBibitem
\bibitem[H{\"a}fner and M{\"u}ller(2026)H{\"a}fner, and M{\"u}ller]{ref29}
H{\"a}fner,~G.; M{\"u}ller,~M. {GPU}-Accelerated Continuum Dynamics of Block Copolymer Blends and Solutions. \emph{J. Chem. Phys.} \textbf{2026}, \emph{164}, 024906\relax
\mciteBstWouldAddEndPuncttrue
\mciteSetBstMidEndSepPunct{\mcitedefaultmidpunct}
{\mcitedefaultendpunct}{\mcitedefaultseppunct}\relax
\EndOfBibitem
\bibitem[McMullen(1993)]{ref15}
McMullen,~W.~E. A Fourth-Order, Density-Functional, Random-Phase Approximation Study of Monomer Segregation in Phase-Separated, Lamellar, Diblock-Copolymer Melts. \emph{Macromolecules} \textbf{1993}, \emph{26}, 1027--1036\relax
\mciteBstWouldAddEndPuncttrue
\mciteSetBstMidEndSepPunct{\mcitedefaultmidpunct}
{\mcitedefaultendpunct}{\mcitedefaultseppunct}\relax
\EndOfBibitem
\bibitem[Wang \latin{et~al.}(2023)Wang, Chen, Tan, and Promislow]{ref16}
Wang,~S.; Chen,~Y.; Tan,~Z.; Promislow,~K. Parameterized Density Functional Models for Block Copolymer Melts. arXiv:2307.00659, 2023; \url{https://arxiv.org/abs/2307.00659}\relax
\mciteBstWouldAddEndPuncttrue
\mciteSetBstMidEndSepPunct{\mcitedefaultmidpunct}
{\mcitedefaultendpunct}{\mcitedefaultseppunct}\relax
\EndOfBibitem
\bibitem[Christlieb \latin{et~al.}(2025)Christlieb, Promislow, Tan, Wang, Wetton, and Wise]{ref30}
Christlieb,~A.; Promislow,~K.; Tan,~Z.; Wang,~S.; Wetton,~B.; Wise,~S.~M. Benchmark Computation of Morphological Complexity in the Functionalized Cahn-Hilliard Gradient Flow. \emph{Commun. Comput. Phys.} \textbf{2025}, \emph{37}, 877--920\relax
\mciteBstWouldAddEndPuncttrue
\mciteSetBstMidEndSepPunct{\mcitedefaultmidpunct}
{\mcitedefaultendpunct}{\mcitedefaultseppunct}\relax
\EndOfBibitem
\bibitem[Fredrickson(2006)]{ref17}
Fredrickson,~G.~H. \emph{The Equilibrium Theory of Inhomogeneous Polymers}; Oxford University Press: New York, 2006\relax
\mciteBstWouldAddEndPuncttrue
\mciteSetBstMidEndSepPunct{\mcitedefaultmidpunct}
{\mcitedefaultendpunct}{\mcitedefaultseppunct}\relax
\EndOfBibitem
\bibitem[Semenov(1985)]{ref18}
Semenov,~A.~N. Contribution to the Theory of Microphase Layering in Block-Copolymer Melts. \emph{Sov. Phys. JETP} \textbf{1985}, \emph{61}, 733--742\relax
\mciteBstWouldAddEndPuncttrue
\mciteSetBstMidEndSepPunct{\mcitedefaultmidpunct}
{\mcitedefaultendpunct}{\mcitedefaultseppunct}\relax
\EndOfBibitem
\bibitem[Tomiyoshi \latin{et~al.}(2024)Tomiyoshi, Honda, Kawakatsu, Murashima, Uehara, and Deguchi]{ref19}
Tomiyoshi,~Y.; Honda,~T.; Kawakatsu,~T.; Murashima,~T.; Uehara,~E.; Deguchi,~T. Density Functional Theory for Cyclic Block Copolymer Melts. \emph{Macromolecules} \textbf{2024}, \emph{57}, 10704--10716\relax
\mciteBstWouldAddEndPuncttrue
\mciteSetBstMidEndSepPunct{\mcitedefaultmidpunct}
{\mcitedefaultendpunct}{\mcitedefaultseppunct}\relax
\EndOfBibitem
\bibitem[Barrat \latin{et~al.}(2005)Barrat, Fredrickson, and Sides]{ref20}
Barrat,~J.-L.; Fredrickson,~G.~H.; Sides,~S.~W. Introducing Variable Cell Shape Methods in Field Theory Simulations of Polymers. \emph{J. Phys. Chem. B} \textbf{2005}, \emph{109}, 6694--6700\relax
\mciteBstWouldAddEndPuncttrue
\mciteSetBstMidEndSepPunct{\mcitedefaultmidpunct}
{\mcitedefaultendpunct}{\mcitedefaultseppunct}\relax
\EndOfBibitem
\bibitem[Cochran \latin{et~al.}(2006)Cochran, Garcia-Cervera, and Fredrickson]{ref22}
Cochran,~E.~W.; Garcia-Cervera,~C.~J.; Fredrickson,~G.~H. Stability of the Gyroid Phase in Diblock Copolymers at Strong Segregation. \emph{Macromolecules} \textbf{2006}, \emph{39}, 2449--2451\relax
\mciteBstWouldAddEndPuncttrue
\mciteSetBstMidEndSepPunct{\mcitedefaultmidpunct}
{\mcitedefaultendpunct}{\mcitedefaultseppunct}\relax
\EndOfBibitem
\bibitem[Dimitriyev \latin{et~al.}(2023)Dimitriyev, Reddy, and Grason]{ref23}
Dimitriyev,~M.~S.; Reddy,~A.; Grason,~G.~M. Medial Packing, Frustration, and Competing Network Phases in Strongly Segregated Block Copolymers. \emph{Macromolecules} \textbf{2023}, \emph{56}, 7184--7202\relax
\mciteBstWouldAddEndPuncttrue
\mciteSetBstMidEndSepPunct{\mcitedefaultmidpunct}
{\mcitedefaultendpunct}{\mcitedefaultseppunct}\relax
\EndOfBibitem
\bibitem[Ceniceros and Fredrickson(2004)Ceniceros, and Fredrickson]{ref24}
Ceniceros,~H.~D.; Fredrickson,~G.~H. Numerical Solution of Polymer Self-Consistent Field Theory. \emph{Multiscale Model. Simul.} \textbf{2004}, \emph{2}, 452--474\relax
\mciteBstWouldAddEndPuncttrue
\mciteSetBstMidEndSepPunct{\mcitedefaultmidpunct}
{\mcitedefaultendpunct}{\mcitedefaultseppunct}\relax
\EndOfBibitem
\end{mcitethebibliography}

\end{document}